\documentclass[sigconf]{acmart}

\usepackage{graphicx}
\usepackage{url}
\usepackage{color}
\usepackage{paralist}
\usepackage{fancyvrb}
\usepackage{booktabs} 
\usepackage{makecell}
\usepackage{algorithmicx}
\usepackage{algorithm}
\usepackage[noend]{algpseudocode}
\usepackage{amsmath}
\usepackage{listings}

\newcommand{\new}[1]{{\color{black}#1}}

\long\def\comment#1{}   
\newcommand{\q}[1]{\lq\lq{}{}#1\rq\rq{}{}}

\newcommand{\MethA}{$M_{\neg p}$}
\newcommand{\MethB}{$M_{p}$}
\newcommand{\MethC}{$M_{pj}$}
\newcommand{\MethD}{$M_{pjf}$}

\algrenewcommand\alglinenumber[1]{\scriptsize #1:}

\renewcommand\footnotetextcopyrightpermission[1]{} 
\begin{document}

\title[Estimating the Cost of Executing Link Traversal based SPARQL Queries]{Estimating the Cost of Executing \\Link Traversal based SPARQL Queries}

\author{Antonis Sklavos}
\email{antonis@sklavos.io}
\affiliation{%
  \institution{Information Systems Laboratory, ICS-FORTH \& University of Crete}
  \city{Heraklion}
  \country{Greece}
}

\author{Yannis Tzitzikas}
\email{tzitzik@ics.forth.gr}
\orcid{0000-0001-8847-2130}
\affiliation{%
  \institution{Information Systems Laboratory, ICS-FORTH \& University of Crete}
  \city{Heraklion}
  \country{Greece}
}

\author{Pavlos Fafalios}
\email{fafalios@ics.forth.gr}
\orcid{0000-0003-2788-526X}
\affiliation{%
  \institution{Information Systems Laboratory, ICS-FORTH}
  \city{Heraklion}
  \country{Greece}
}

\begin{abstract}
An increasing number of organisations in almost all fields have started adopting semantic web technologies for publishing their data as open, linked and interoperable (RDF) datasets, queryable through the SPARQL language and protocol. 
\textit{Link traversal} has emerged as a SPARQL query processing method that exploits the \textit{Linked Data} principles and the dynamic nature of the Web to dynamically discover data relevant for answering a query by resolving online resources (URIs) during query evaluation. 
However, the execution time of link traversal queries can become prohibitively high for certain query types due to the high number of resources that need to be accessed during query execution. 
In this paper we propose and evaluate \new{baseline} methods for estimating the evaluation cost of link traversal queries. 
Such methods can be very useful for deciding on-the-fly the query execution strategy to follow for a given query, thereby reducing the load of a SPARQL endpoint and increasing the overall reliability of the query service. 
To evaluate the performance of the proposed methods, we have created (and make publicly available) a ground truth dataset consisting of 2,425 queries. 
\end{abstract}



\maketitle

\pagestyle{plain} 

\section{Introduction}
In the last years, a constantly increasing body of knowledge is made available on the web in the RDF format~\cite{manola2004rdf}, including cross-domain knowledge bases, such as DBpedia~\cite{lehmann2015dbpedia} and   Wikidata~\cite{vrandevcic2014wikidata}, but also domain-specific datasets, such as  ClaimsKG~\cite{tchechmedjiev2019claimskg} (fact-checking), ORKG~\cite{jaradeh2019open} (scholarly communication),  
DrugBank~\cite{wishart2008drugbank,wishart2018drugbank} (drugs),
Semantic Layers~\cite{fafalios2017building} (web archives),
Sampo portals~\cite{hyvonen2022digital} (digital humanities).
In general, semantic technologies are increasingly used in a plethora of topical domains for making data available openly and interoperable for research and wide use \cite{schmachtenberg2014adoption}.
Following the \textit{Linked Data} principles\footnote{\url{https://www.w3.org/DesignIssues/LinkedData.html}}~\cite{heath2011linked}, such online RDF datasets can be directly accessed and queried by interested parties and external applications.

SPARQL 
is currently the standard language and protocol for querying and manipulating RDF datasets. 
However, the low reliability of SPARQL endpoints is the major bottleneck that deters the exploitation of these knowledge bases by real applications~\cite{buil2013sparql,debattista2018evaluating}. 
For instance, \cite{buil2013sparql}~tested 427 public endpoints and found that their performance can vary by up to 3-4 orders of magnitude, while only 32.2\% of public
endpoints can be expected to have monthly up-times of 99-100\%. The more recent work in~\cite{debattista2018evaluating} confirmed these performance and reliability issues, showing also that over
a period of 6 months, at least 11\% of the considered endpoints became less reliable.

Link traversal~\cite{hartig2013overview,umbrich2015link} is an alternative SPARQL query processing method which relies on the \textit{Linked Data} principles to answer a query by accessing (resolving) online web resources (URIs) dynamically, during query execution, without accessing endpoints. 
This query processing method is based on robust web protocols (HTTP, IRI), is in line with the dynamic nature of the Web, motivates decentralisation, and enables answering queries without requiring data providers to setup and maintain costly servers/endpoints.
However, the execution time of link traversal queries can become prohibitively high for certain types of queries due to the very high number of resources that need to be resolved at query execution time for retrieving their RDF triples~\cite{fafalios2019many}. This performance issue is a reason that deters the wider adoption of this query evaluation method. 

In this paper, we focus on this problem and study methods to estimate the execution cost of queries that can be answered through link traversal. Our focus is, in particular, on \textit{zero-knowledge} link traversal, which does not consider a starting graph or seed URIs for initiating the traversal, relying only on URIs that exist in the query pattern or that are dynamically retrieved during query execution.\footnote{According to \cite{fafalios2019many}, more than 85\% of the queries submitted to five popular endpoints are answerable through zero-knowledge link traversal.}
We consider as \textit{execution cost} the number of URIs that need to be accessed and resolved in real time because this affects both the query execution time and the amount of data that need to be transferred over the network, being at the same time  independent of the underlying link traversal implementation/engine. 
By estimating this cost, a query service can decide on the fly the query execution strategy to follow for an incoming SPARQL query, based on factors such as the expected query execution time of link traversal and the availability (or current load) of the endpoint.

\begin{figure}[h]
    \vspace{-1mm}
    \centering
    \fbox{\includegraphics[width=8cm]{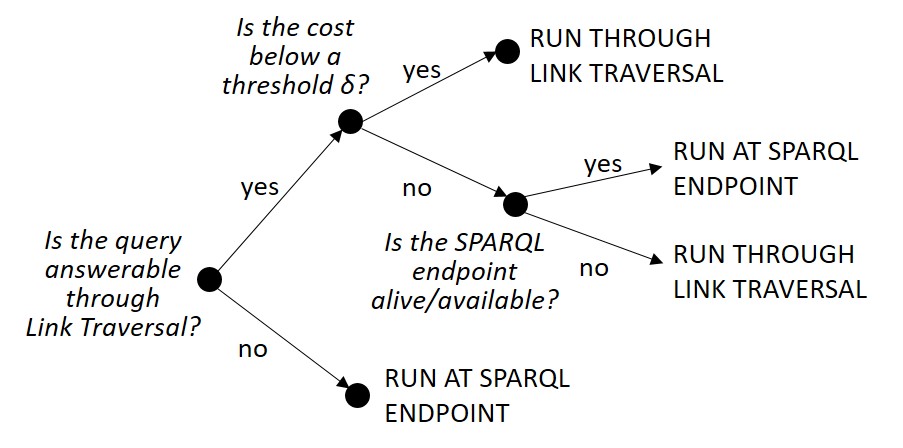}}
    \vspace{-3mm}
    \caption{Deciding the query execution strategy to follow for answering a SPARQL query.}
    \label{fig:decisionTree}
    \vspace{-1mm}
\end{figure}

Fig.~\ref{fig:decisionTree} shows a decision tree that can be considered by a SPARQL query service for deciding (in real time) on the query execution strategy to follow, aiming at improving the overall reliability of the query service without  significantly affecting its response times.
If the incoming query is answerable through zero-knowledge link traversal and the estimated execution cost is low, then the query will bypass the SPARQL endpoint (or the federation of the endpoints in case of a distributed environment) and will be executed through link traversal. 
If the cost is high, the service can check the availability of the endpoint(s), e.g., by running an ASK query. In case of availability, the query will be executed at the endpoint(s). Otherwise, the query will run though link traversal since it is preferable to get a delayed response than getting no results at all.  
Note here that the costs of checking the answerability of a query through link traversal, computing the link traversal cost, and checking the availability of a SPARQL endpoint, are negligible (a few ms).
For endpoints that receive a high number of queries, such a query execution plan can highly reduce their load, and thus improve their overall reliability, since a large number of queries (of low link traversal cost) will bypass the endpoint(s) and be evaluated through link traversal using the robust HTTP protocol.

To enable the comparative evaluation of cost estimation methods, we build and make publicly available a ground truth dataset consisting of \new{2,425} queries. 
Using this ground truth, we experimentally evaluate four cost estimation methods that consider different aspects of the query, such as the type of predicates or the appearance of joins. The results showed that, considering  predicate statistics, which can be computed easily (and only once) in a pre-processing step, together with joins and FILTER clauses, provides the best cost estimation performance. 
 
In a nutshell, in this paper we make the following contributions:

\begin{itemize}
    \item We study the main factors that affect the cost of executing a  SPARQL query through zero-knowledge link traversal and propose a set of baseline methods to estimate it.
    \item We provide a ground truth dataset for the problem per se.
    \item We evaluate the performance of the baseline methods using the introduced ground truth dataset  and a use case over a cross-domain knowledge base, in particular DBpedia~\cite{lehmann2015dbpedia}.
\end{itemize}

The implementation of the proposed methods and the ground truth dataset are publicly available.\footnote{\label{foot1}\url{https://github.com/isl/LDAQ-CostEstimators}}

The remainder of this paper is organised as follows:
Sect.~\ref{sec:rw} provides the required background and describes related work. 
Sect.~\ref{sec:approach} details the main characteristics that affect the link traversal cost and introduces four baseline methods to estimate the cost.
Sect.~\ref{sec:experiments} introduces the ground truth dataset and presents evaluation results.
Finally, Sect.~\ref{sec:conclusion} concludes the paper and discusses directions for future work.

\section{Background and Related Work}
\label{sec:rw}

\subsection{Background}

Link traversal is a SPARQL query execution method that accesses online resources in real time (during query evaluation) in order to retrieve the data needed for answering a SPARQL query~\cite{hartig2013overview,umbrich2015link}. 

Consider, for instance, the below query that is to be executed over the DBpedia knowledge base:
\small
\begin{Verbatim}[frame=lines,numbers=left,numbersep=1pt]
 PREFIX dbr: <http://dbpedia.org/resource/> 
 PREFIX dbo: <http://dbpedia.org/ontology/> 
 SELECT ?birthDate  WHERE {
   dbr:Nelson_Mandela dbo:birthDate ?birthDate }
\end{Verbatim}
\normalsize
The query contains one triple pattern requesting the birth date of Nelson Mandela. 
If we run the query over the SPARQL endpoint of DBpedia, we get back the literal \q{1918-07-18} (xsd:date).  
To answer the query, a link traversal query execution method first needs to access the dereferenceable URI of Nelson Mandela (\url{https://dbpedia.org/resource/Nelson_Mandela}) for retrieving the RDF triples contained in this resource, and then evaluate the triple pattern over these RDF triples. 
In this example, only one resource needs to be accessed during query execution.\footnote{We consider that the URI of a predicate of a triple pattern is not dereferenced during link traversal because it (usually) does not contain all triples containing the particular URI as predicate, thus it does not help binding the subject or object variable.}

Consider now the below query which requests all persons that influenced Plato together with a description of each of them:
\small
\begin{Verbatim}[frame=lines,numbers=left,numbersep=1pt]
 PREFIX dbr: <http://dbpedia.org/resource/> 
 PREFIX dbo: <http://dbpedia.org/ontology/> 
 SELECT ?influencer ?influencerDescription  WHERE {
   dbr:Plato dbo:influencedBy ?influencer . 
   ?influencer dbo:abstract ?influencerDescription 
         FILTER (lang(?influencerDescription) = 'en') }
\end{Verbatim}
\normalsize
In this case, link traversal first needs to access the URI of Plato in order to find the persons that influenced him, and then the URIs of all Plato's influencers for retrieving their description (14 influencers, according to DBpedia). Thus, link traversal needs to access 15 resources in total for answering this query.

A limitation of this query execution method is that not all queries are answerable through \textit{zero-knowledge} link traversal. For example, the below query selects all triples whose object is a URI:
\small
\begin{Verbatim}[frame=lines,numbers=left,numbersep=1pt]
 SELECT * WHERE {
   ?subject ?predicate ?object FILTER isURI(?object) }
\end{Verbatim}
\normalsize
This query cannot be answered without considering a starting graph or seed URIs, since there is no starting point (e.g., a URI in the query) that can be used for initiating the link traversal. 

Our work focuses on queries that are answerable through zero-knowledge link traversal. The analysis of query logs in \cite{fafalios2019many} has shown that such queries correspond to the majority ($>$85\%) of the queries submitted to known endpoints. 

An interesting way to directly run queries through zero-knowledge link traversal is by using SPARQL-LD~\cite{fafalios2015sparql,fafaliosquerying}, a generalisation of SPARQL 1.1 which
extends the applicability of the SERVICE operator to enable querying any HTTP web source containing RDF data,
like dereferencable URIs, online RDF files, or web pages embedded with RDFa or JSON-LD.\footnote{\url{https://github.com/fafalios/sparql-ld}} 
The below query is an example of SPARQL-LD query that retrieves all persons that influenced Plato and their descriptions, without needing to access DBpedia's endpoint.
The query first accesses the URI of Plato for retrieving his influencers (lines 4-5), and then queries the bound URI of each influencer for getting the description (lines 6-8):
\small
\begin{Verbatim}[frame=lines,numbers=left,numbersep=1pt]
 PREFIX dbr: <http://dbpedia.org/resource/> 
 PREFIX dbo: <http://dbpedia.org/ontology/> 
 SELECT ?influencer ?influencerDescription  WHERE {
   SERVICE dbr:Plato { 
      dbr:Plato dbo:influencedBy ?influencer } 
   SERVICE ?influencer { 
      ?influencer dbo:abstract ?influencerDescription 
            FILTER (lang(?influencerDescription) = 'en') } }
\end{Verbatim}
\normalsize


Since link traversal might need access to a large number of remote resources for answering a query (e.g., thousands of URIs), making query execution time prohibitively high, in this paper we provide methods that can estimate the query execution cost of a SPARQL query before its execution. 
Studying query execution strategies that exploit cost estimation (like the decision tree in Fig.~\ref{fig:decisionTree}) is out of the scope of this paper.

\subsection{Related Work}

We first review the related literature on link traversal and then position our work.  

\subsubsection{Link Traversal Approaches}

Link Traversal exploits the \textit{Linked Data} principles~\cite{heath2011linked} to dynamically discover data relevant for answering a SPARQL query~\cite{umbrich2015link}. 

The approaches in \cite{hartig2009executing,hartig2012sparql} and \cite{miranker2012diamond} follow RDF links by resolving URIs that exist in the query and in partial results. The URIs are resolved over the HTTP protocol into RDF data which is continuously added to the queried dataset using an iterator-based pipeline. 
\cite{hartig2011zero} studies how the evaluation order in link traversal affects the size of the results and the query execution cost, and proposes a heuristics-based method to optimise query execution. 
\cite{bouquet2009querying}, \cite{hartig2012sparql} and \cite{harth2012completeness} discuss the notion of {\em completeness} and propose semantics to restrict the range of link traversal queries.
\cite{umbrich2015link} studies the effectiveness of link traversal-based query execution and proposes reasoning extensions to help finding additional answers.

Another direction of work on link traversal relies on pre-built indexes for finding sources to look up during query execution \cite{harth2010data,tian2011enhancing,wagner2012top}.
These approaches can determine all potentially relevant URIs at the beginning of query execution, which enables to fully parallelize the data retrieval process. However, there is a cost of initialising and maintaining the indexes.

The works in \cite{fafalios2019many} and \cite{fafalios2019answering} focus on queries that can be answered directly on the live Web of Data, without considering a starting graph or seed URIs for initiating the link traversal. This \textit{zero-knowledge} approach corresponds to the {\em query-reachable} completeness class as introduced in \cite{harth2012completeness}. The starting point of link traversal in this case is one or more URIs that exist in the query's graph pattern while additional URIs are resolved only if this is needed for binding the variables of a triple pattern. 
The same works provide open source methods for i) examining the answerability of a query through zero-knowledge link traversal, and ii) transforming an answerable query to a SPARQL-LD query that is executed through this query evaluation method.\footnote{\url{https://github.com/fafalios/LDaQ}}

\subsubsection{Positioning}

We focus on link traversal queries that can be answered directly on the live Web of Data, without considering a starting graph or seed URIs (zero-knowledge link traversal~\cite{fafalios2019answering} and query-reachable completeness class~\cite{harth2012completeness}).

Our work extends \cite{fafalios2019answering} by 
(a)~providing baseline methods to estimate the execution time of a query pattern that is answerable through zero-knowledge link traversal, 
(b)~providing a ground truth dataset for the problem per se, and (c)~evaluating the performance of the proposed methods using the introduced dataset and a use case over DBpedia.


Note here that, estimating the link traversal cost complements works on \textit{query optimisation} \cite{stocker2008sparql, hartig2011zero,huang2011estimating,tsialiamanis2012heuristics,yannakis2018heur}, such as selectivity-based query reordering methods~\cite{hartig2011zero,yannakis2018heur}. Query optimisation tackles a related but different problem. For instance, a query executor can apply query pattern reordering before estimating the link traversal cost, or avoid applying optimisation if the estimated cost is low.
Ideally, the query executor first applies query optimisation for defining the more efficient join order and then it computes the cost of zero-knowledge link traversal considering this fixed join order.

\section{Estimating the Query Cost}
\label{sec:approach}

We consider as \textit{query cost} the number of remote resources that have to be accessed and retrieved during query execution. This number affects both the query evaluation time as well as the amount of data that is transferred through the network, and is independent of the underlying implementation/engine of zero-knowledge link traversal.

Another factor that affects the query evaluation performance is the size of the remote resources, i.e., the number of triples contained in these resources. However, in a dynamic web context it is impossible to know in advance the number of triples contained in a remote resource without first accessing and retrieving it. Since we aim at estimating the cost of a query before its execution, this factor is not considered in our proposed methods. 

Below, we first discuss the most common query characteristics that affect the link traversal cost (Sect.~\ref{subsec:costChars}) and then introduce four methods to estimate the cost (Sect.~\ref{subsec:methods}).

\subsection{Characteristics affecting the query cost}
\label{subsec:costChars}

\subsubsection{Characteristic \#1}
Given a basic graph pattern, the first characteristic that affects the query cost is the number of distinct URIs that appear as subjects or objects in the graph pattern's triples. These URIs are resolved during query evaluation for either binding variables or making the necessary joins. 

\subsubsection{Characteristic \#2}
The second characteristic is the number of distinct variables whose bindings can bind other variables. For each such variable, the query needs to resolve all its URI bindings for evaluating the graph pattern, which can be very costly for cases of large number of bindings. We call these variables {\em necessary-to-resolve} variables.

\begin{definition}[Necessary-to-resolve variable]
\label{def:def1}
A necessary-to-resolve variable in a query graph pattern is a variable whose URI bindings need to be resolved for binding another variable in the same query graph pattern. 
\end{definition}

Consider, for example, the below graph pattern which requests all authors together with the venues of their publications:
\small
\begin{Verbatim}[frame=lines,numbers=left,numbersep=1pt]
 ?author a :Author .
 ?author :hasPublication ?publication .
 ?publication :inVenue ?venue  
\end{Verbatim}
\normalsize
The pattern first has to access the URI of the {\tt :Author} class for binding the variable {\tt ?author}.
Then, it needs to access the URI of each author for binding the variable {\tt ?publication}. Thus, the variable {\tt ?author} is a necessary-to-resolve variable.
Finally, the pattern needs to access the URI of each author publication for binding the variable {\tt ?venue}. Thus, {\tt ?publication} is another necessary to resolve variable, while the variable {\tt ?venue} is not since there is no need to resolve it. 

If we arbitrarily consider that the number of authors returned by the {\tt :Author} URI is 10,000 and that each author has 50 publications, then the \textit{maximum} number of URIs that need to be accessed for evaluating the query is 510,001 (1 for binding {\tt ?author} + 10,000 for binding the variable {\tt ?publication} of each author + 10,000$\times$50 for binding the variable {\tt ?venue} of each author publication). 

\new{Note here that there might be several common URIs in the bindings of one or more necessary-to-resolve variables (e.g., publications shared by multiple authors, in our example). Since it is impossible to know the values of the bindings without executing the query, we consider the \textit{worst-case} scenario in which all bindings are different.}

\subsubsection{Characteristic \#3}
The third characteristic that affects the query cost is the type and value of the predicate used to bind a necessary-to-resolve variable. 
First, if the predicate is a variable, then the number of bindings of the necessary-to-resolve variable can be very high since all different predicates connecting the subject with the object are considered. For example, the below query pattern requests all URI properties of authors (i.e., their related entities) together with their labels: 
\small
\begin{Verbatim}[frame=lines,numbers=left,numbersep=1pt]
 ?author a :Author .
 ?author ?property ?relatedEntity .
 ?relatedEntity :label ?label 
\end{Verbatim}
\normalsize
The number of bindings of the necessary-to-resolve variable {\tt ?re\-la\-tedEntity} (which needs to be resolved for binding {\tt ?label}) can be very high since all properties of an author are considered.

If now the predicate is a URI, its value can affect the number of bindings of the corresponding necessary-to-resolve variable. There are predicates for which the objects have, on average, small number of subjects but also predicates for which the objects can have a large number of subjects. For instance, we know that the capital of a country can be only one, thus for the {\tt :hasCapital} predicate we expect one \textit{object binding}. On the contrary, we can expect a relatively large number of \textit{object bindings} for the {\tt :hasMember} predicate (depending on the context).
Likewise, we can expect a small number of \textit{subject bindings} for some predicates and a large number for some other. For instance, on average we expect a large number of \textit{subject bindings} for the predicate {\tt :birthPlace} (there are many persons with the same birth place), but a small number for the predicate {\tt :hasCapital} (there is only one country for a given capital city).

\subsubsection{Characteristic \#4}
The fourth characteristic is the number of star-shaped joins that limit the bindings of necessary-to-resolve variables (which, in this case, are the common variables in the joins). Consider, for example, the below query:
\small
\begin{Verbatim}[frame=lines,numbers=left,numbersep=1pt]
 ?author a :Author .
 ?author :directorOf ?institution .
 ?author :hasPublication ?publication .
 ?publication :inVenue ?venue  
\end{Verbatim}
\normalsize
The second triple pattern limits the bindings of the variable {\tt ?author} of the first triple pattern to only those having a {\tt :directorOf} property (forming a star-shaped join).
In this case, the number of bindings of the variable {\tt ?author} can be highly reduced before moving to the third pattern. 
For example, if from the 10,000 authors in the knowledge base only 100 are directors, then the maximum number of URIs that need to be accessed for evaluating the query is highly limited from 510,001 to 15,001 (1 for binding the variable {\tt ?author} + 10,000 for binding the variables  {\tt ?institution} and {\tt ?publication} + $100\times50$ for binding the variable {\tt ?venue} of each author publication, if we roughly consider that each author has 50 publications). 

The number of bindings can be also reduced if the necessary-to-resolve variable is involved in a chain-shaped join whose other subject or object elements are URIs or literals.
For example, consider the below query pattern which instead of having the {\tt :directorOf} property, it has a triple pattern (as first one) which requires  the authors to belong to a specific party: 
\small
\begin{Verbatim}[frame=lines,numbers=left,numbersep=1pt]
 :party12 :hasMember ?author . 
 ?author a :Author .
 ?author :hasPublication ?publication .
 ?publication :inVenue ?venue  
\end{Verbatim}
\normalsize
The two triples in lines 1 and 2 form a chain-shaped join which can highly limit the number of bindings of the variable {\tt ?author}.

\subsubsection{Characteristic \#5}
Another query characteristic that can highly limit the bindings of a necessary-to-resolve variable and thus the query cost is the presence and position of FILTER clauses. The FILTER clause is a constraint which restricts the solutions (variable bindings) of the whole group of triple patterns in which it appears. Consider, for example, the below query pattern:
\vspace{0.5mm}
\small
\begin{Verbatim}[frame=lines,numbers=left,numbersep=1pt]
 ?author a :Author .
 ?author :directorOf ?institution . 
 ?author :birthDate ?birthDate FILTER(year(?birthDate)>1985)
 ?author :hasPublication ?publication .
 ?publication :inVenue ?venue }
\end{Verbatim}
\normalsize
The pattern is the same with that of the first example of \textit{Characteristic \#4} with an addition of one triple pattern and one FILTER operator (line 3). The triple pattern requests the birth date of each author while the filter operator restricts the accepted values of the {\tt ?birthDate} bindings to only those of year after 1985. Since the {\tt ?author} variable exists in the same triple with the {\tt ?birthDate} variable, the filter operator can highly limit its bindings. For instance, if the number of authors who are directors and have a birth date after 1985 is 10, then the maximum number of URIs that need to be accessed is further limited to 10,511 (1 for binding the variable {\tt ?author} + 10,000 for checking the {\tt directorOf} property of all authors + 10 for binding the variable {\tt ?publication} of each author who is director and has a birth date after 1985 + 10$\times$50 for binding the variable {\tt venue} of each author publication).

\subsubsection{Characteristic \#6}
A last characteristic that can potentially  affect the query cost is the order of the triples and FILTERs in the graph pattern. Query writing is not always optimal and this can affect the query execution time if the underlying SPARQL implementation does not apply an optimisation technique, e.g., a query re-ordering method \cite{yannakis2018heur}. 
For instance, in the query pattern above (example of Characteristic \#5), if we move the second and third triples to the end and the SPARQL implementation does not apply any pattern re-ordering method, the query cost is highly increased because the query needs to first retrieve the venues of all publications of all 10,000 authors before restricting the bindings of the {\tt ?author} variable. 
For being widely applicable and implementation independent, we do not require that a specific query optimisation method must be applied before estimating the query cost.

\subsection{Cost Estimation Methods}
\label{subsec:methods}

Considering the above-mentioned characteristics that can affect the query execution cost of link traversal, we now provide baseline methods to estimate it. The implementation of all methods is publicly available on GitHub (see Footnote \ref{foot1}).

\subsubsection{Method 1 - Predicates agnostic (\MethA)}

Our first baseline method considers the number of URIs in the query (Characteristic \#1) and also tries to estimate the expected number of bindings of each \textit{necessary-to-resolve} variable (Characteristics \#2 and \#3). For the latter, we consider very limited knowledge about the underlying knowledge base. In particular, we consider constant values for the below five parameters:
\begin{enumerate}
    \item Average number of entity \textit{outgoing} properties
    \item Average number of entity \textit{incoming} properties
    \item Average number of \textit{subject bindings} for any given property (except \textit{rdf:type}) and object
    \item Average number of \textit{subject bindings} for the property \textit{rdf:type} and a given object/class (i.e., average number of instances per class)
    \item Average number of object bindings for a given subject and property
\end{enumerate}

One can easily compute these values in a pre-processing step (and only once), e.g., by running SPARQL queries or accessing the RDF dumps. Listings  \ref{q:param1query}-\ref{q:param5query} provide the queries for each of the five parameters.

\begin{figure}
\small
\begin{lstlisting}[caption={SPARQL query for computing the entities' average number of outgoing properties (Parameter \#1).},captionpos=b,frame=lines,label={q:param1query}]
SELECT (AVG(?count) AS ?average)
WHERE {
  { SELECT ?x (COUNT(DISTINCT ?y) AS ?count) 
    WHERE {
      ?x a ?type . ?x ?y ?z } GROUP BY ?x }} 
\end{lstlisting}

\vspace{1.5mm}

\begin{lstlisting}[caption={SPARQL query for computing the entities' average number of incoming properties (Parameter \#2).},captionpos=b,frame=lines,label={q:param2query}]
SELECT (AVG(?count) AS ?average)
WHERE {
  { SELECT ?z (COUNT(DISTINCT ?y) AS ?count) 
    WHERE {
      ?x ?y ?z . ?z a ?type } GROUP BY ?z }} 
\end{lstlisting}

\vspace{1.5mm}

\begin{lstlisting}[caption={SPARQL query for computing the average number of subject bindings for any property except \textit{rdf:type} (Parameter \#3).},captionpos=b,frame=lines,label={q:param3query}]
SELECT (AVG(?count) AS ?average)
WHERE {
 { SELECT ?z (COUNT(DISTINCT ?x) AS ?count) 
   WHERE {
    ?x ?y ?z FILTER (?y!=rdf:type) } GROUP BY ?z } } 
  \end{lstlisting}

\vspace{1.5mm}

\begin{lstlisting}[caption={SPARQL query for computing the average number of instances per class (Parameter \#4).},captionpos=b,frame=lines,label={q:param4query}]
SELECT (AVG(?count) AS ?average)
WHERE {
  { SELECT ?z (COUNT(DISTINCT ?x) AS ?count) 
    WHERE { ?x a ?z } GROUP BY ?z } } 
\end{lstlisting}

\vspace{1.5mm}

\begin{lstlisting}[caption={SPARQL query for computing the average number of object bindings for any property (Parameter \#5).},captionpos=b,frame=lines,label={q:param5query}]
SELECT (AVG(?count) AS ?average)
WHERE {
  { SELECT ?x (COUNT(DISTINCT ?z) AS ?count) 
    WHERE { ?x ?y ?z } GROUP BY ?x } } 
\end{lstlisting}
\normalsize
\vspace{-5mm}
\end{figure}

\subsubsection{Method 2 - Predicates aware (\MethB)} 

This method extends the first method by considering the actual value (URI) of each predicate in the query pattern (Characteristic~\#3). In particular, we can pre-compute (only once, in a pre-processing step) the \textit{average number of subject bindings} (see query in Listing~\ref{q:sparqlAvGObjBindings}) and the \textit{average number of object bindings}  (see query in Listing.~\ref{q:sparqlAvGSubjBindings}) for a large number of frequent predicates or for the full list of predicates used by the underlying knowledge base(s).
For example, the predicate \url{http://dbpedia.org/ontology/genre} (the genre of a thing, e.g., of a music group, film, etc.) has average number of object bindings 1.8 (an object has, on average, around 2 genres) and average number of subject bindings 56.9 (there are, on average, around 57 objects having the same genre). 
These pre-computed numbers are then exploited in real time for estimating the number of bindings of the necessary-to-resolve variables contained in the query pattern.
If the query pattern contains an unknown predicate, then we consider an average value as in \MethA. 

Although the actual predicate selectivity can be skewed for a particular subject or object (since, for example, for the same predicate there might be subjects with very high number of object bindings and subjects with very low number), it provides a good estimate that is adequate in our (cost estimation) use case since it can distinguish the cases of always-small  and always-large number of subject/object bindings.

\begin{figure}
\small
\begin{lstlisting}[caption={SPARQL query for computing the average number of object bindings for a particular predicate.},captionpos=b,frame=lines,label={q:sparqlAvGObjBindings}]
SELECT (AVG(?count) AS ?average)
WHERE {
  { SELECT ?x (COUNT(DISTINCT ?z) AS ?count) 
    WHERE { ?x <predicate> ?z } GROUP BY ?x } } 
\end{lstlisting}

\vspace{1.0mm}

\begin{lstlisting}[caption={SPARQL query for computing the average number of subject bindings for a particular predicate.},captionpos=b,frame=lines,label={q:sparqlAvGSubjBindings}]
SELECT (AVG(?count) AS ?average)
WHERE {
  { SELECT ?z (COUNT(DISTINCT ?x) AS ?count) 
    WHERE { ?x <predicate> ?z } GROUP BY ?z } } 
\end{lstlisting}
\vspace{-6mm}
\end{figure}

\subsubsection{Method 3 - Predicates+Joins aware (\MethC)}
Here we extend \MethB\ by considering the star-shaped joins of necessary-to-resolve variables (Characteristic~\#4). 
In particular, if a necessary-to-resolve variable participates in a star-shaped join, we reduce its estimated number of bindings by a constant factor $f_1$ (which can take a value in the range [0.0, 1.0]). For instance, if $f_1=0.5$ and the estimated number of bindings of a necessary-to-resolve variable is 500, then this is reduced to 250 (0.5*500). 

To decide on the value of $f_1$, we consider a set of training queries whose real cost is known (more in Sect. \ref{sec:experiments}).
Moreover, we do not consider the joins of the first and last query triple patterns since they do not affect the number of bindings of necessary-to-resolve variables (their bindings must be resolved regardless of whether they participate in star-shaped joins or not).

\subsubsection{Method 4 - Predicates+Joins+Filters aware (\MethD)}
Our last baseline method extends the previous method (\MethC) by also considering the appearance of FILTER clauses in the query graph pattern (Characteristic \#5). 
In particular, similar to the case of joins, if a necessary-to-resolve variable exists in a FILTER clause and there is a triple pattern after that FILTER expression which requires resolving the URI bindings of the necessary-to-resolve variable, we reduce its estimated number of bindings by a constant factor $f_2$. 

To decide on the value of $f_2$, we again consider a set of training queries for which we know their real cost (more in Sect. \ref{sec:experiments}).

\section{Ground Truth Dataset \& Evaluation}
\label{sec:experiments}

We first describe the ground truth dataset we created for enabling the evaluation of cost estimation methods (Sect.~\ref{subsec:GT}). Then, we evaluate the four baseline methods described in the previous section (Sect.~\ref{subsec:evalSetup}-\ref{subsec:evalResults}). Finally, we summarise the main findings (Sect.~\ref{subsec:execSum}).

\subsection{Ground Truth}
\label{subsec:GT}
The use of any cross-domain knowledge base (or federation of knowledge bases) providing resolvable URIs is adequate for the objective of our evaluation (examining the performance of cost estimation methods); we only need a diverse set of queries that are answerable through zero-knowledge link traversal. 

To this end, we built a ground truth dataset by using \textit{real} query logs of DBpedia provided by the USEWOD series of workshops \cite{luczak2016usewod}. 
In particular, we gathered a set of distinct queries and computed their real link traversal cost. To compute the cost, we first transformed them to SPARQL-LD \cite{fafaliosquerying} queries using the algorithm provided in~\cite{fafalios2019many} and  without applying any pattern reordering method.  Then, we executed the SPARQL-LD queries and counted the number of remote resources that each query needed to access for providing the results through zero-knowledge link traversal.\footnote{Any implementation of zero-knowledge link traversal that does not apply a query optimisation technique is expected to provide the same real cost. } 

To build the dataset, we discarded queries that are not answerable through zero-knowledge link traversal (using the algorithm provided in \cite{fafalios2019many}), as well as a large number of queries having \textit{cost~=~1}, i.e., queries that require accessing a single URI that appears in the query pattern (by excluding them we can preform a more representative evaluation of the introduced methods).  
Also, we did not consider queries that make use of the {\tt UNION} keyword or property paths (handing such cases is part of our future work), as well as queries with errors or that timed out.

The final dataset consists of 2,425 queries (see Footnote \ref{foot1} for an access link). 
For each query we provide:
\begin{itemize}
 \item[i)] the transformed {\tt SPARQL-LD} query that executes the query pattern through zero-knowledge link traversal
 \item[ii)] the real query cost of link traversal
 \item[iii)] all URIs that had to be accessed by the link traversal, together with the date and time we run the query 
 \item[iv)] the Notation3 (N3) files containing the triples of all the accessed URIs (57,138 unique files in total for all queries)
\end{itemize}

\subsection{Evaluation Setup}
\label{subsec:evalSetup}
We evaluated the four methods described in the previous section (\MethA, \MethB, \MethC, \MethD) using the introduced ground truth dataset. 
First, we split the dataset randomly into two equal parts and used the one part (\textit{train}) for optimising the factors $f_1$ and $f_2$ of methods 3 and 4, respectively, and the other part (\textit{test}) for evaluating  performance on unseen queries. The considered value for both $f_1$ and $f_2$ is 0.9.
For the parameters of \MethA, we considered the below constant values (decided by running SPARQL queries on DBpedia): 
\begin{enumerate}
    \item Average number of entity \textit{outgoing} properties = 25
    \item Average number of entity \textit{incoming} properties = 5
    \item Average number of \textit{subject bindings} for any given (no \textit{rdf:type}) property and object = 1,505
    \item Average number of \textit{subject bindings} for the property \textit{rdf:type} and a given object/class (i.e., average number of instances per class) = 848
    \item Average number of object bindings for a given subject and property = 1.86
\end{enumerate}

To measure the performance of each method over the set of test queries $Q$ and the considered knowledge base $K$,
we consider the \textit{average absolute difference} (AvgAbsDiff) between the real cost and the estimated cost for all queries in $Q$. In particular: 
\begin{equation*}
    \text{AvgAbsDiff} = \frac{\sum_{q \in Q}{|\text{realCost}(q, K) - \text{estimatedCost}(q, K)|}}{|Q|}
\end{equation*}
We also consider the \textit{percentage difference} of the average real cost compared to the average estimated cost for all test queries (\%AvgDiff). 
This will show us if, on average, the estimated cost is larger or smaller compared to the real cost (and how much). In both measures,  smaller values means better results (closer to the real cost).

\subsection{Evaluation Results}
\label{subsec:evalResults}

Table \ref{tbl:res1} shows the results for the four methods on the full test dataset.
We notice that \MethB, \MethC\ and \MethD\ have a similar performance (from +46\% to +54\% of the real cost), highly outperforming \MethA\ (predicates agnostic). The best performance is achieved by \MethD\ which is predicates-aware and also considers joins and filters. 

\begin{table}[h]
\centering
\caption{Results on full test dataset.}
\label{tbl:res1}
\vspace{-2mm}
\begin{tabular}{lrr}
\toprule
Method & AvgAbsDiff & ~~~\%AvgDiff \\
\midrule
\MethA\ & 523.9 & +1,037\%  \\ 
\MethB\ & 100.9  & +54\%\\ 
\MethC\ & 98.5 & +49\% \\ 
\MethD\ & 97.8  & +46\% \\
\bottomrule
\end{tabular}
\vspace{-2mm}
\end{table}

To better understand the performance of \MethC\ (Predicate+Joins aware), and its difference compared to \MethB\ (Predicates aware), we now consider only queries from the test dataset that contain star-shaped joins (453 queries, in total).  
Table \ref{tbl:res2} shows the results.
We now see that the performance improvement of \MethC\ compared to \MethB\ is much higher (from 137\% of real cost to 114\%) compared to the results of Table \ref{tbl:res1}, suggesting the positive effect of considering star-shaped joins in cost estimation. 

\begin{table}[h]
\centering
\caption{Results on test dataset considering only queries with star-shaped joins.}
\label{tbl:res2}
\vspace{-2mm}
\begin{tabular}{lrr}
\toprule
Method & AvgAbsDiff & ~~~\%AvgDiff \\
\midrule
\MethA\  & 480.6 & +1,477\%  \\ 
\MethB\  & 91.1  & +137\%\\ 
\MethC\  & 84.6 & +114\% \\ 
\MethD\  & 84.5  & +113\% \\
\bottomrule
\end{tabular}
\vspace{-2mm}
\end{table}

We do the same considering only queries that contain both star-shaped joins and filters (436 queries, in total), in order to examine the performance of \MethD\ (Predicates+Joins+Filters aware) compared to \MethC\ (Predicates+Joins aware).
Table \ref{tbl:res3} shows the results.
We now see that the performance of these two methods is almost the same, with \MethD\ slightly outperforming \MethC. 
This very small difference is justified by the small number of queries in our test dataset that contain star-shaped joins but no FILTER (453$-$436 = 17 queries, in total), 
as well as by the fact that a high percentage of queries in out test dataset (94\%) contain the filter clause at the very end of their pattern (thus, not affecting the bindings of a necessary-to-resolve variable). 

\begin{table}[h]
\centering
\caption{Results on test dataset considering only queries with star-shaped joins \textit{and} filters.}
\label{tbl:res3}
\vspace{-2mm}
\begin{tabular}{lrr}
\toprule
Method & AvgAbsDiff & ~~~\%AvgDiff\\
\midrule
\MethA\  & 429.0 & +1,340\%  \\ 
\MethB\ & 76.2 & +93\%\\ 
\MethC\  & 71.2 & +75\% \\ 
\MethD\   & 71.2  & +74\% \\
\bottomrule
\end{tabular}
\vspace{-2mm}
\end{table}

In general, we notice that the estimated cost is higher than the real cost in all cases. This is an expected result because, as described in Sect.~\ref{sec:approach}, we consider the worst-case scenario in which all bindings of the necessary-to-resolve variables are different (it is impossible to know the values of the bindings without first executing the query).

\subsection{Executive Summary}
\label{subsec:execSum}
The evaluation results demonstrate that pre-computing information about the predicates used in a knowledge base can highly improve the cost estimation performance, as expected. As we have seen, it is straightforward to pre-compute this information (e.g., by running SPARQL queries or accessing RDF dumps), and is something that needs to be done only once. 

We also noticed that considering joins and FILTER clauses through constant reduction factors does not highly improve cost estimation compared to the predicates-aware method, as one would probably expect. This is a good motivation for studying more fine-grained methods of exploiting joins and filters that do not consider constant factors, e.g., through training based on specific patterns.

The four described baseline methods are generic, i.e., they can work over any knowledge base or federation of knowledge bases. We only need to pre-compute the information required by the predicates aware methods, while if such information is not available for a predicate we can use a constant (average) value. 

Finally, we have defined \textit{cost} as the number of remote resources that need to be accessed and retrieved during query execution. Here, multiple such resources can be downloaded in parallel, or one can use a caching mechanism for frequent resources or a pattern reordering method, which means that the \textit{real} query execution performance can be significantly improved. Optimising the query execution of link traversal is a different problem and out of the scope of this paper.

\section{Conclusions}
\label{sec:conclusion}

We have analyzed the main query characteristics that affect the cost of executing SPARQL queries through \textit{zero-knowledge link traversal}, a query execution method that relies on robust web protocols (HTTP, IRI) and the dynamic nature of the web for answering a query by accessing online resources during query evaluation. 

Based on these query characteristics, we introduced four baseline methods to estimate the link traversal cost. 
The first method considers very limited knowledge about the underlying knowledge base, the second considers predicate statistics (average number of subject/object bindings, computed once in a pre-processing step), the third method extends the second by also considering star-shaped joins, while the last method extends all previous methods by also considering FILTER clauses. 

Such query cost estimation methods can be very useful for deciding on-the-fly the query execution strategy to follow, based on factors such as the availability and load of the SPARQL endpoint(s) at query time, as well as the estimated cost of each considered query execution method. The aim is to improve the overall reliability of the query service without significantly affecting its response times.

To measure the performance of the cost estimation methods, we have created (and make publicly available; cf. Footnote \ref{foot1}) a benchmark comprising DBpedia queries that are answerable through zero-knowledge link traversal. 
The experiments over this ground truth have shown that, considering predicate statistics together with joins and FILTER clauses provides the best cost estimation (around $46\%$ higher than the real cost, on average).

In the future, we plan to study additional, more fine-grained cost estimation methods, e.g., one that do not consider constant factors in case of joins and FILTERs.
Another direction for future work includes the implementation of query execution strategies that exploit cost estimation for deciding on-the-fly  on the query evaluation method to follow. 

\section*{Acknowledgements}
This work has received funding from the European Union's Horizon 2020 research and innovation programme under the Marie Sklodowska-Curie grant agreement No 890861 (Project \q{ReKnow}).

\bibliographystyle{ACM-Reference-Format}
\balance
\bibliography{MAIN} 

\end{document}